\documentclass[a4paper,12pt]{article}
\usepackage[utf8]{inputenc}
\usepackage{cancel}
\usepackage{tikz}
\usepackage{ulem}
\usepackage{amsfonts}
\usepackage{amssymb}
\usepackage{graphicx}
\usepackage{amsmath}
\usepackage{enumerate}
\usepackage{mathtools}
\usepackage{subfig}
\usepackage{color}
\usepackage{tikz}
\usepackage{float}
\usepackage{here}
\usepackage{cite}
\usepackage{mathrsfs}
\usepackage{float,epsfig}
\usepackage{dcolumn}
\usepackage{graphicx}
\usepackage{bm}
\usepackage{amsmath,amssymb,amsthm}
\usepackage[colorlinks=true,linkcolor=blue,citecolor=red]{hyperref}
\usepackage{multirow}
\usepackage[toc,page]{appendix}

\usetikzlibrary{arrows.meta}
\usetikzlibrary{bending}
\usetikzlibrary{calc}
\newcommand{\be}{\begin{equation}}
\newcommand{\ee}{\end{equation}}
\newcommand{\bea}{\setlength\arraycolsep{2pt} \begin{eqnarray}}
\newcommand{\eea}{\end{eqnarray}}

\def\0{{\sst{(0)}}}
\def\1{{\sst{(1)}}}
\def\2{{\sst{(2)}}}
\def\3{{\sst{(3)}}}
\def\4{{\sst{(4)}}}
\def\5{{\sst{(5)}}}
\def\6{{\sst{(6)}}}
\def\7{{\sst{(7)}}}
\def\8{{\sst{(8)}}}
\def\sst#1{{\scriptscriptstyle #1}}

\makeatletter \@addtoreset{equation}{section}

\definecolor{lime}{HTML}{A6CE39}

\begin{document}

\title{{\normalsize \textbf{\Large Probing Quantum Entanglement from Quantum
Correction to Newtonian Potential Energy}}}
\author{ {\small A. Belhaj$^1$\thanks{%
a-belhaj@um5r.ac.ma}, S. E. Ennadifi$^{2}$\thanks{%
ennadifis@gmail.com}, L. Jebli$^{3}$\thanks{%
larbi.jebli@gmail.com} \thanks{\textbf{Authors in alphabetical order.}} 
\hspace*{-8pt}} \\
{\small $^1$D\'{e}partement de Physique, \'Equipe des Sciences de la
mati\`ere et du rayonnement, ESMaR}\\
{\small Facult\'e des Sciences, Universit\'e Mohammed V de Rabat, Rabat,
Morocco} \\
{\small $^2$LHEP-MS, Facult\'e des Sciences, Universit\'e Mohammed V de
Rabat, Rabat, Morocco } \\
{\small $^{3}$LPTHE, Dep. of Physics, Faculty of Sciences, Ibn Zohr
University, PO Box 8106, Agadir, Morocco}}
\maketitle

\begin{abstract}
Inspired by string theory ideas, we probe quantum entanglement from the
gravitational potential energy. Concretely, we reconsider the study of
quantum corrections to the Newtonian potential energy by treating a massive
two-particle system $m_{1}$ and $m_{2}$ with size dimensions $r_{1}$ ad $%
r_{2}$ where the two particles separated by a distance $d$ are under only
their mutual classical gravitational interaction $V_{r}\left( r_{1}\text{, }%
r_{2}\right) $. Exploring such a size-dependent gravitational behavior and
taking the limit $r_{1}$, $r_{2}\ll d$, we investigate the associated
quantum biparticle state and express its evolution after an interaction time 
$\tau $. Among others, we show that the two masses cannot be separable due
to the induced gravitational entanglement in terms of the accumulated
quantum phase $\delta \phi =\delta V_{g}\tau /\hbar $. By analogy with the
classical gravity, we derive the expression of the resulting extremely weak
entanglement force from the corresponding gravitational entanglement energy.
Then, we provide certain entanglement diagnostics.

\end{abstract}

\newpage

\section{Introduction}

It is now believed that the basic underlying framework for the physics of
the Universe is quantum mechanics (QM). The foundation for quantum field
theory that describes successfully the physics of elementary particles and
their governing forces within the Standard Model of Particle Physics by help
of the gauge theory formalism \cite{1,2,3,4,5,6,7,8,9}. One of the source of
theoretical interest in probing such a foundation is quantum entanglement
(QE) \cite{1,10,11,12,13,14,15}. Although such a non-classical correlation
has been perceived in a borad diversity of systems and length scales
extending from the microscopic to the macroscopic levels \cite{146,147}, up
to present, however, entanglement has stayed for the most part unprobed at
the high energies attainable at particle colliders, mainly LHC. Recently, a
probe of QE effects via top quark-antitop quark pairs and their decays in
the proton--proton collision events at a center-of-mass energy of $13$ $TeV$
recorded with the ATLAS detector has been reported \cite{148,149,150}. This
has unlocked new roads to examine the fundamental QM properties.

In some QM interpretations such as decoherence theory, QE is considered as a
more fundamental property \cite{10,11}. In particular, it has been argued
that an observable QE between massive objects could be generated by their
mutual gravitational interaction through a significant induced phase \cite%
{16,17,18}. In recent decades, it is believed that QE is related to the
spacetime structure \cite{17,18}. Although such a connection with the
emergence of spacetime is inspiring it remains indirect. In modern physics,
the spacetime dynamics and its interaction with matter has been understood
within general relativity and governed by the Einstein field equations \cite%
{19,20}. While the weakness of gravity, which is one of the most universal
of all interactions, has made it difficult to test theories on its nature it
remains a key to interpret the observed data in numerous experiments \cite%
{21,22,23,24,25}. Using a classical gravity theory, a meaningful
understanding of quantum observational data is possible at the cost of
including additional specific parameters \cite{26,27,28,29,151}. Though
there is no experimental evidence of quantum features of gravity, it is
strongly suggested that it should also be treated within the quantum
framework after the successful unification of electromagnetic, weak and
strong interactions \cite{1,2,3,4,5,6,7,8,9}. Trying to fit gravity into the
QM framework has been one of the great challenges in theoretical and modern
physics over the past decades \cite{30}. The most famous example concerns
string theory. The latter describes the dynamics of one-dimensional objects
called strings. The internal dimension of such objects have been exploited
to provide gravitons in the quantum spectrum of closed string configurations 
\cite{300,301,302}.

Inspired by string theory ideas, considering the point-like
particles of ordinary physics as one-dimensional objects with internal
dimensions, we investigate a possible link between QE and the gravitational
field by means of the potential energy. Precisely, we deal with a massive
two-particle system $m_{1}$ and $m_{2}$ with the size dimensions $r_{1}$ ad $%
r_{2}$ where the two particles separated by a distance $d$ are under only
their mutual classical gravitational interaction $V_{r}\left( r_{1}\text{, }%
r_{2}\right) $. By considering such a size-dependent gravitational behavior
and taking $r_{1,2}\ll d$, we provide a quantum correction to the Newtonian
potential energy. Then, we study the associated quantum biparticle state and
express its evolution after an interaction time $\tau $. More precisely, we
show that the two massive particles cannot be separable due to the induced
gravitational entanglement in terms of an accumulated quantum phase $\delta
\phi =\delta V_{g}\tau /\hbar $. By analogy with the classical gravity, we
derive the expression of the resulting extremely weak entanglement force
from the corresponding gravitational entanglement energy. Finally, we
discuss certain entanglement diagnostics.

The organization of this work is as follows. In section 2, we reconsider the
study of quantum corrections to the Newtonian potential energy. In section
3, we investigate the probing of QE from quantum corrections to such a
potential energy. In section 4, we approach certain entanglement
diagnostics. The last section is devoted to conclusion and perspectives.

\section{Quantum correction to the gravitational energy}

In the effective field theories framework \cite{1111,1112}, the
quantum effects can be dealt with  and meaningful expectations can be
conducted. Actually, the opener point in an effective field theory is the
dissociation of comprehended physics at the working scales from
uncomprehended one at very higher energy scales. As we climb to higher
energies/shorter distances, experiments have revealed that new degrees of
freedom and new interactions arise and get in scene. We have no support to
doubt that the effects of  the  present theory of gravity are the entire
scenario at the highest energies. Without managing ungroundless suppositions
about what is taking place at high energies, predictions at current energies
are allowed in an effective theory approach.  In doing so, the emerging divergences at high energies arising from loop diagrams (on account of the unreliable high energy parts of the loop integration) are not predictions of the effective theory, and thus do not anyhow get into any physical outcomes, thing which is not the case for quantum effects of the low energy part of the theory being the quantum predictions of the effective theory at low energies.  Accordingly, at
very short distance scales,  the classical description of gravitation is no
longer correctly usable since it should start to experience quantum effects
in that limit. An expository employment of this approach is the computation
of quantum effects on the gravitational interaction of two massive particles
where the power-law quantum corrections to the ordinary $r^{-1}$
potential are computable.

Now, with all this in mind, we reconsider the study of quantum
corrections to the gravitation theory by means of the potential energy which
is firmly only approximately correct. Indeed, in the case of either
considerable masses or velocities there are verified relativistic
corrections that must be taken into account according to the framework of
the general relativity theory \cite{151}. Analogously, at very short
distance scales, we would also anticipate that quantum corrections have not
to be ignored and thus would lead to a modification in the gravitational
potential, as it was the case for the modified Coulomb interaction by the
radiative corrections of quantum electrodynamic \cite{152}.

To start, we consider a three dimensional system described by two massive
particles 1 and 2 with masses $m_{1}$ and $m_{2}$, respectively. These two
massive particles are separated by a distance $d$ from each other. The
associated gravitational potential energy reads as 
\begin{equation}
V_{0}=-\frac{Gm_{1}m_{2}}{d}  \label{eq0}
\end{equation}%
where $G$ is the Newton constant. An examination shows that this energy form
is an approximated expression since it has been also exploited for
non-punctual configurations. However, such compact geometries should involve
geometric parameters controlling either the size or the shape associated
with the internal dimension. This concept has been explored in
string theory that considers the point-like particles of ordinary physics,
including quantum field theory, as one-dimensional compact objects with
internal sizes \cite{300,301,302}. Developments in such a theory has shown
the existence of higher dimensional objects called branes. Many geometrical
configurations of branes have been dealt with such as toric and spherical
ones with various parameters controlling the involved size and shape
behaviors \cite{303,304}. Inspired by objects with internal sizes, we could
modify the above equation by considering a spherical configuration. It
turns out that this geometric configuration is the simple one exhibiting
only one real size parameter being the radius. In this spirit, the above
energy equation should carry data on such a geometric parameter. Roughly, we
study the setting where the two particles have spherical geometries of radii 
$r_{1}$ and $r_{2}$ assumed to be extremely small compared to the separation
distance $d$. They are supposed to be only under the mutual gravitational
interaction generated by their masses. In this setup, the above potential
energy for such a mutual gravitational interaction should be refined as
follows 
\begin{equation}
V_{r}=-\frac{Gm_{1}m_{2}}{d+\left( \delta r_{1}+\delta r_{2}\right) }.
\label{eq1}
\end{equation}%
This extra quantity $\delta r_{1}+\delta r_{2}$ characterizes now the
possible dependence of the potential energy on the size parameters by
modifying the separation distance $d$. Assuming that the radii $r_{1}$ and $%
r_{2}$ are extremely small compared to the separation distance $d$ ($\delta
r_{1,2}\ll d$), the gravitational potential energy (\ref{eq1}) could be
expanded as 
\begin{equation}
V_{r}\left( r_{1},r_{2}\right) \simeq -\frac{Gm_{1}m_{2}}{d}\left( 1-\frac{%
\left( \delta r_{1}+\delta r_{2}\right) }{d}+\frac{\left( \delta
r_{1}+\delta r_{2}\right) ^{2}}{d^{2}}+....\right)   \label{eq2}
\end{equation}%
where the first term, being zeroth one, is the classical mass-centered
gravitational potential. The second one is just a displacement term which
can be absorbed and removed. A close inspection reveals that the higher
order terms generate correlation behaviors which couple the two masses being
proportional to the non-separable terms $\left( \delta r_{1}+\delta
r_{2}\right) ^{n}$. Considering the fact that $d\gg r_{1,2}\ $ and $\left(
\delta r_{1}+\delta r_{2}\right) ^{2}/d^{2}\gg \left( \delta r_{1}+\delta
r_{2}\right) ^{n}/d^{n}$ for $n>2$, the significant apparent contributions
responsible for such a correlation start to appear from the third order
term. For simplicity reasons, we restrict ourselves to such a term where the
above potential energy (\ref{eq2}) reduces to the following relevant form 
\begin{equation}
V_{r}\left( r_{1},r_{2}\right) \simeq -\frac{Gm_{1}m_{2}}{d}\left( 1+\frac{%
\left( \delta r_{1}+\delta r_{2}\right) ^{2}}{d^{2}}+\mathcal{O}((\delta
r_{1}+\delta r_{2})^{3})\right) .  \label{eq3}
\end{equation}%
In what follows, we show that the size quadratic term could induce a quantum
correction to the classical Newtonian potential energy from the extended
expression given in terms of the particle sizes. A way to do is to view the
distance $\delta r_{1}+\delta r_{2}$ as an uncertainty about the particle
positions. Indeed, the possible displacement of their centers of mass either
to the right $\delta r_{1,2}^{+}$ or the left $\delta r_{1,2}^{-}$ from
equilibrium so as the radii \ $r_{1}$ and $r_{2}$ of the two particles read
now as 
\begin{equation}
r_{1,2}^{+}\rightarrow r_{1,2}+\delta r_{1,2}^{+},\text{ \ \ }\delta
r_{1,2}^{-}+r_{1,2}\text{ }  \leftarrow r_{1,2}^{-}.\text{\ }  \label{eq4}
\end{equation}%
In this way, the studied system can be interpreted as one-dimensional
harmonic oscillators of masses $m_{1}$ and $m_{2}$ with vibrational
frequencies $\omega _{1}$ and $\omega _{2}$, respectively. Based on this
identification, we obtain the quantum expressions for the small displacement
as 
\begin{equation}
\delta r_{1}\sim \sqrt{\frac{\hbar }{m_{1}\omega _{1}}}\text{, \ \ \ }\delta
r_{2}\sim \sqrt{\frac{\hbar }{m_{2}\omega _{2}}}  \label{eq7}
\end{equation}%
where $\hbar $ is the reduced Planck constant. Using (\ref{eq3}) and (\ref%
{eq7}), we get the quantum corrected potential energy 
\begin{equation}
V_{g}\left( \omega _{1},\omega _{2}\right) \simeq -\frac{Gm_{1}m_{2}}{d}%
\left[ 1+\frac{\hbar }{d^{2}}\left( \frac{1}{m_{1}\omega _{1}}+\frac{1}{%
m_{2}\omega _{2}}+\frac{2}{\sqrt{m_{1}m_{2}\omega _{1}\omega _{2}}}\right) +%
\mathcal{O}(\hbar ^{2})\right] .  \label{eq8}
\end{equation}%
The first term which does not involve any power of $\hbar $ describes the
standard classical potential. However, the second term is a relevant quantum
correction provided by a linear contribution of $\hbar $ being 
\begin{equation}
\delta V_{g}\left( \omega _{1},\omega _{2}\right) \simeq -\frac{\hbar
Gm_{1}m_{2}}{d^{3}}\left( \frac{1}{m_{1}\omega _{1}}+\frac{1}{m_{2}\omega
_{2}}+\frac{2}{\sqrt{m_{1}m_{2}\omega _{1}\omega _{2}}})\right) .
\label{eq9}
\end{equation}%
This highly suppressed $d^{-3}$ new correction is a correlation term since
it contains the mixed term $\left( m_{1}m_{2}\omega _{1}\omega _{2}\right)
^{-1}$. At this level, we would like to add certain comments on such a
quantum correction to the gravitational potential energy. First, this
finding could be considered as an alternative way to generate a quantum
effect, with respect to the one reported in \cite{151}. The present
correction could be supported by string theory scientific revolution where
the internal dimension of physical objects could predict a quantum version
of the gravity. Precisely, the closed string theory, assuming that such
objects are circles, contains gravitons in the associated quantum spectrum 
\cite{300,301,302}. The quantum effect provided here will be considered in
what follows in probing the induced gravitational entanglement between the
two-particles.

\section{Probe of entanglement from gravity}

Having pictured the state of the two separated masses in a vibrational
motion, a qubit can be encoded in their internal degrees of freedom. In
fact, we can now think of each mass as an effective positional qubit with
its two states being the spatial states $\left\vert r^{+}\right\rangle $ and 
$\left\vert r^{-}\right\rangle $. Assuming the states of the two particles
as the localized Gaussian wave-packets whose the widths are much less than
their radii $r_{1}$ and $r_{2}$, we can write the following orthogonality
relations 
\begin{equation}
\left\langle r_{i}^{a}\mid r_{j}^{b}\right\rangle =\delta _{b}^{a}\delta
_{j}^{i}
\end{equation}%
where one has used $a,b=+,-$ and $i,j=1,2$. For the masses, the initial
normalized states $\left\vert \psi _{1}\left( t=0\right) \right\rangle $ and 
$\left\vert \psi _{2}\left( t=0\right) \right\rangle $ in the Hilbert spaces 
$H_{1}^{r_{1}\text{ }}$and $H_{2}^{r_{2}\text{ }}$ of the two particles,
which are held adjacently in the superposition of spatially localized states
due to the uncertainty about the positions of the centers of masses, could
be written as follows 
\begin{eqnarray}
\left\vert \psi _{1}\left( t=0\right) \right\rangle &=&\frac{1}{\sqrt{2}}%
\left( \left\vert r_{1}^{+}\right\rangle +\left\vert r_{1}^{-}\right\rangle
\right)  \label{eq10} \\
\left\vert \psi _{2}\left( t=0\right) \right\rangle &=&\frac{1}{\sqrt{2}}%
\left( \left\vert r_{2}^{+}\right\rangle +\left\vert r_{2}^{-}\right\rangle
\right) .  \label{eq11}
\end{eqnarray}%
The whole state $\left\vert \psi _{12}\left( t=0\right) \right\rangle $ of
the bipartite system in the Hilbert space 
\begin{equation}
{H}_{12}^{r\text{ }}=H_{1}^{r_{1}\text{ }}\otimes H_{2}^{r_{2}\text{ }}
\label{eq12}
\end{equation}%
takes the form 
\begin{eqnarray}
\left\vert \psi _{12}\left( t=0\right) \right\rangle &=&\left\vert \psi
_{1}\left( t=0\right) \right\rangle \otimes \left\vert \psi _{2}\left(
t=0\right) \right\rangle  \label{eq13} \\
&=&\frac{1}{2}\left( \left\vert r_{1}^{+}r_{2}^{+}\right\rangle +\left\vert
r_{1}^{+}r_{2}^{-}\right\rangle +\left\vert r_{1}^{-}r_{2}^{+}\right\rangle
+\left\vert r_{1}^{-}r_{2}^{-}\right\rangle \right) .  \notag
\end{eqnarray}%
After a certain interaction time $t=\tau $, the evolution of the overall
state (\ref{eq13}) under the gravitational interaction can be driven by the
Shr\"{o}dinger evolution equation, without the kinetic term, 
\begin{equation}
\widehat{V}_{g}\left\vert \psi _{12}\left( \tau \right) \right\rangle
=i\hbar \partial _{\tau }\left\vert \psi _{12}\left( \tau \right)
\right\rangle  \label{eq14}
\end{equation}%
where $\widehat{V}_{g}$ represents the operator for the gravitational
potential energy. According to the depiction of the particle positions given
in (\ref{eq4}) in terms of their respective small right-left displacements $%
\delta r_{1,2}^{+,-}$, we can describe the corresponding gravitational
potential operator. Actually, there are four possible displacements such as 
\begin{eqnarray}
r_{1}^{+} &\rightarrow &r_{1}+\delta r_{1}^{+},\text{ \ \ }%
r_{2}^{+}\rightarrow r_{2}+\delta r_{2}^{+} \\
\text{ \ \ }\delta r_{1}^{-}+r_{1}\text{ } &  \leftarrow &r_{1}^{-},\text{
\ \ \ \ }\delta r_{2}^{-}+r_{2}\text{ }  \leftarrow r_{2}^{-}
\end{eqnarray}%
for right-right $r_{1}^{+}r_{2}^{+}$ and left-left $r_{1}^{-}r_{2}^{-}$
displacements respectively, and 
\begin{eqnarray}
r_{1}^{+} &\rightarrow &r_{1}+\delta r_{1}^{+},\text{ \ \ }\delta
r_{2}^{-}+r_{2}\text{ }  \leftarrow r_{2}^{-} \\
\text{ \ \ }\delta r_{1}^{-}+r_{1}\text{ } &  \leftarrow&r_{1}^{-},\text{%
\ \ \ }r_{2}^{+}\rightarrow r_{2}+\delta r_{2}^{+}
\end{eqnarray}%
for right-left $r_{1}^{+}r_{2}^{-}$ and left-right $r_{1}^{-}r_{2}^{+}$
displacements, respectively. For the first case where the displacement of
the two particles is in the same direction, the characterizing distance of
the particle positions remains the same. For the seconde case where the
displacement of the two particles is in the opposite directions, however,
the characterizing distance of the particle positions on which the potential
energy depends is extended. Based on these reasonings, we provide a matrix
representation of the operator $\widehat{V}_{g}$. Roughly, the corresponding
potential energy operator in the basis $\left\{ \left\vert
r_{1}^{+}r_{2}^{+}\right\rangle ,\left\vert r_{1}^{+}r_{2}^{-}\right\rangle
,\left\vert r_{1}^{-}r_{2}^{+}\right\rangle ,\left\vert
r_{1}^{-}r_{2}^{-}\right\rangle \right\} $ can be taken as 
\begin{equation}
\widehat{V}_{g}=\left( 
\begin{array}{cccc}
V_{g}-\delta V_{g} & 0 & 0 & 0 \\ 
0 & \delta V_{g} & 0 & 0 \\ 
0 & 0 & \delta V_{g} & 0 \\ 
0 & 0 & 0 & V_{g}-\delta V_{g}%
\end{array}%
\right) .
\end{equation}%
The solution of (\ref{eq14}) gives rise to a gravitational-induced quantum
phase $\phi= \phi \left( V_{g}\right) $. In particular, we find 
\begin{equation}
\left\vert \psi \left( \tau \right) \right\rangle _{12}=\frac{1}{2}\left(
e^{-i\phi ^{++}}\left\vert r_{1}^{+}r_{2}^{+}\right\rangle +e^{-i\phi
^{+-}}\left\vert r_{1}^{+}r_{2}^{-}\right\rangle +e^{-i\phi ^{-+}}\left\vert
r_{1}^{-}r_{2}^{+}\right\rangle +e^{-i\phi ^{--}}\left\vert
r_{1}^{-}r_{2}^{-}\right\rangle \right)  \label{eq16}
\end{equation}%
where the induced phases $\phi ^{++},\phi ^{+-},\phi ^{-+}$ and $\phi ^{--}$
have, according to the gravitational potential (\ref{eq8}) and (\ref{eq9}),
the following forms 
\begin{eqnarray}
\phi ^{++} &=&\phi ^{--}\equiv \phi =\frac{\left( V_{g}-\delta V_{g}\right)
\tau }{\hbar },  \label{eq17} \\
\phi ^{+-} &=&\phi ^{-+}\equiv \phi ^{\prime }=\frac{V_{g}\tau }{\hbar }.
\label{eq18}
\end{eqnarray}%
Taking such phases, the evolved whole state (\ref{eq16}) can be expressed as
follows 
\begin{equation*}
\left\vert \psi \left( \tau \right) \right\rangle _{12}=\frac{e^{-i\phi }}{2}%
\left[ \left( \left\vert r_{1}^{+}r_{2}^{+}\right\rangle +\left\vert
r_{1}^{-}r_{2}^{-}\right\rangle \right) +e^{-i\delta \phi }\left( \left\vert
r_{1}^{+}r_{2}^{-}\right\rangle +\left\vert r_{1}^{-}r_{2}^{+}\right\rangle
\right) \right]
\end{equation*}%
where $\delta \phi $ is considered as a quantum induced term given by 
\begin{equation}
\delta \phi =\delta V_{g}\tau /\hbar .  \label{eq20}
\end{equation}%
Using such a phase, the evolved state reads now as 
\begin{equation}
\left\vert \psi _{12}\left( \tau \right) \right\rangle =\frac{e^{-i\frac{%
\left( V_{g}-\delta V_{g}\right) \tau }{\hbar }}}{2}\left[ \left( \left\vert
r_{1}^{+}r_{2}^{+}\right\rangle +\left\vert r_{1}^{-}r_{2}^{-}\right\rangle
\right) +e^{-i\frac{\delta V_{g}\tau }{\hbar }}\left( \left\vert
r_{1}^{+}r_{2}^{-}\right\rangle +\left\vert r_{1}^{-}r_{2}^{+}\right\rangle
\right) \right] .  \label{eq21}
\end{equation}%
Obviously, the state (\ref{eq21}) is entangled since it is not factorisable.
More precisely, the induced entanglement between the two masses has its
origin in the quantum accumulated phase given in (\ref{eq20}). This is
guaranteed only for the non-zero phase shifts by requiring the condition 
\begin{equation}
\delta \phi \neq 2\pi n  \label{25}
\end{equation}%
where $n$ is a positive integer. This entanglement increases monotonically
over $\delta \phi $ given by 
\begin{equation}
\delta \phi =\frac{Gm_{1}m_{2}\tau }{d^{3}}\left( \frac{1}{m_{1}\omega _{1}}+%
\frac{1}{m_{2}\omega _{2}}+\frac{2}{\sqrt{m_{1}m_{2}\omega _{1}\omega _{2}}}%
\right) \neq 2\pi n  \label{eq23}
\end{equation}%
which is in general not equal to $2\pi n$ if proper parameters $d$ and $%
\omega _{1,2}$ are chosen. Thus, $\left\vert \psi _{12}\left( t=\tau \right)
\right\rangle $ can not be factorized and the entanglement between the two
test masses can be created. The result shows that the gravitational
interaction can induce quantum entanglement between two masses, and
consequently confirms that gravity might behave as a quantum entity if we
recognize our size-dependent potential energy approach considered in (\ref%
{eq1}).

Going further, we probe this QE between the spatial states of the two
massses. This could prove that a quantum-like gravity force is behind QE
between them as a classical force cannot entangle the states of spatially
separated objects. Concretely, by analogy with the classical gravity, we can
derive a resulting QE force from the entanglement energy which corresponds
here to the quantum correction to the potential energy given in (\ref{eq9}) 
\begin{equation}
E_{entg}\equiv \delta V_{QC}.  \label{eq28}
\end{equation}%
Therefore, the QE force could be simply expressed as 
\begin{equation}
F_{entg}\simeq \frac{\hbar Gm_{1}m_{2}}{d^{3}}\left[ \frac{1}{m_{1}\omega
_{1}^{2}}+\frac{1}{m_{1}\omega _{2}^{2}}+\frac{1}{\sqrt{m_{1}m_{2}}}\left( 
\frac{1}{\sqrt{\omega _{1}^{3}\omega _{2}}}+\frac{1}{\sqrt{\omega _{1}\omega
_{2}^{3}}}\right) \right] .
\end{equation}
Although this QE force is a weak Newton-like force going as $\sim \hbar
d^{-3}$, it can guarantee entangled behavior of the whole system. What we
would like to show here is the relation between gravity felt by the test
particles and the induced QE of the whole system characterized by the weak
entanglement force. Maybe the issue is that the corresponding interaction
potential energy cannot be written as two independent terms with regard to
the position degree of freedoms. In this case, the evolution equation (\ref%
{eq14}) can be separated into two independent equations and admits a
solution in the form of a product state of the initial states\footnote{%
Such an induced entanglement shows that QE can be also generated through
other similar interactions, for instance Coulomb interaction, but we think
it is not universal as gravity since it deals only with electromagnetically
charged particles.}. Thus, the origin of QE strongly depends on the geometry
properties of the interacting fields. It is reasonable to argue that there
is a strong connection between QE and the structure of spacetime.

Having established a possible bridging scenario between gravity and QM, we
move now to consider certain entanglement measures.

\section{Entanglement measure}

In this section, we would like to discuss some QE quantities. It has been
remarked that there are various quantitive measures of such a quantum
behavior. For the sake of simplicities, we limit our examination to two
quantities. The first one is the purity of the reduced density matrices
being obtained from the following density matrix 
\begin{equation}
\rho \left( \tau \right) =\left\vert \psi _{12}\left( \tau \right)
\right\rangle \left\langle \psi _{12}\left( \tau \right) \right\vert.
\end{equation}%
In the basis $\left\{ \left\vert r_{1}^{+}r_{2}^{+}\right\rangle ,\left\vert
r_{1}^{+}r_{2}^{-}\right\rangle ,\left\vert r_{1}^{-}r_{2}^{+}\right\rangle
,\left\vert r_{1}^{-}r_{2}^{-}\right\rangle \right\} $, it has been found to
be

\begin{equation}
\rho \left( \tau \right) =\frac{1}{4}\left( 
\begin{array}{cccc}
1 & e^{i\delta \phi } & e^{i\delta \phi } & 1 \\ 
e^{-i\delta \phi } & 1 & 1 & e^{-i\delta \phi } \\ 
e^{-i\delta \phi } & 1 & 1 & e^{-i\delta \phi } \\ 
1 & e^{i\delta \phi } & e^{i\delta \phi } & 1%
\end{array}%
\right) .  \label{eq24}
\end{equation}%
This density matrix $\rho $ can be quantified by measuring the quantum
purity $\mathcal{P}\left( \rho \right) $ as

\begin{equation}
\mathcal{P}\left( \rho \right) =Tr \left( \rho ^{2}\right)  \label{eq25}
\end{equation}%
which attains its maximal value $1$ for pure states and its minimal value of 
$\mathcal{P}\left( \rho \right) =\frac{1}{d}$ for maximally mixed state for $%
d$-dimensional systems. To approach such a quantity for the two subsystems $%
1 $ and $2$, we need to determine the corresponding reduced density
matrices. Straightforward calculations provide 
\begin{equation}
\rho _{1}\left( \tau \right) =\rho _{2}\left( \tau \right) =\frac{1}{2}%
\left( 
\begin{array}{cc}
1 & \cos (\delta \phi ) \\ 
\cos (\delta \phi ) & 1%
\end{array}%
\right) .  \label{eq26}
\end{equation}%
Having these reduced density matrices, we can now calculate their purities
through 
\begin{equation}
Tr \left( \rho _{1}^{2}\left( \tau \right) \right) =Tr \left( \rho
_{2}^{2}\left( \tau \right) \right) =\frac{1+\cos (\delta \phi )^{2}}{2}.
\label{eq27}
\end{equation}%
Quite remarkably, the purities of the subsystems $1$ and $2$ are equal.
Thus, we conclude that the purity of the a reduced state can be directly
used as a quantifier of QE. The more entangled are the two systems, the more
mixed are their reduced density matrices. According to \cite{18} , this can
be discussed via the quantity 
\begin{equation}
\varepsilon =1-Tr \left( \rho _{1}^{2}\left( \tau \right) \right)
\end{equation}%
In the present case, it has been found to be 
\begin{equation}
\varepsilon =\frac{1-\cos (\delta \phi )^{2}}{2}
\end{equation}%
which depends on the quantum phase $\delta \phi $. Considering this
constraint (\ref{25}), we have $\varepsilon >0$ showing that two subsystems
are entangled. As an alternative, we finally discuss the second measure
relaying on the entropy concept. This diagnostic procedure is called
entanglement entropy. To do so, we should calculate the following quantity 
\begin{equation}
S(\rho _{1})=-Tr(\rho _{1}\log (\rho _{1})).
\end{equation}%
Accordingly, we get 
\begin{equation}
S(\rho _{1})=-\left[ \frac{(1-\cos (\delta \phi ))}{2}\log \left( \frac{%
1-\cos (\delta \phi )}{2}\right) +\frac{(1+\cos (\delta \phi ))}{2}\log
\left( \frac{1+\cos (\delta \phi )}{2}\right) \right] .
\end{equation}%
Using the constraint (\ref{25}), this quantity is positive due the fact that
one has $0\leq \frac{(1\pm \cos (\delta \phi ))}{2}\leq 1$. For $\delta \phi
=(2n+1)\frac{\pi }{2}$, it reaches its maximal value being $S(\rho
_{1})=\log 2$ corresponding to the maximal entanglement behavior.

\section{Conclusion and perspectives}

In this work, though understanding the origin of QE remains a great
challenge, we have attempted to investigate a possible link between QE and
the known fundamental interactions. Concretely, we have established a direct
connection between the gravitational interaction and QE.  Inspired by
string theory paradigm, interpreting the point-like particles of ordinary
physics as one-dimensional objects with internal dimensions, we have
investigated the dependence of the gravitational potential energy, in
addition to the separation distance $d$, on the particle sizes $\sim r_{1,2}$
of the massive two-particle system, where the two particles interact only
through classical gravity. More precisely, by developing the corresponding
potential energy in the limit $d\ll r_{1,2}$ and using an adequate
equivalence, we have expressed the full gravitational potential energy
involving a quantum corrected term with a $\hbar $ linear behavior. After
straightforward calculations dealing with the evolution of the whole quantum
state, we have shown that the ground state of such a system is entangled.
Although the result cannot be practically applied to cosmic systems such as
the Sun-Earth system, its significance comes from the conceptual
implications. Indeed, it confirms and generalizes the idea that the
classical gravitational field can induce QE. Such an entanglement has its
origin in the accumulated non-zero quantum phase $\delta \phi $, by
considering the present size-dependent potential energy approach (\ref{eq1}%
). By analogy with the classical gravity, to probe this QE, we have provided
an entanglement force which has been derived from the the corresponding
energy source, i.e., entanglement energy, identified as the quantum
corrected potential energy (\ref{eq9}). Though this force is found to be
weak $\sim \hbar d^{-3}$ compared to the ordinary Newton force, it can
maintain the whole system entangled. More deeply, since the gravitational
field of the system corresponds to the curvature of spacetime created, it is
reasonable to argue that there is an intrinsic connection between QE and
spacetime geometries. Then, we have discussed certain entanglement
diagnostics. Among others, we have approached the entropy entanglement
supporting the proposed investigation.

In spite of the fact that our work is mostly theoretical and conceptual, we
believe that such an interpretation of gravitational induced QE may provide
some new insights into the nature of gravity, the spacetime geometries and
maybe the arrow of time. Especially, this goes with the current powerful
particle colliders, i.e., LHC exploring fundamental particles and their
governing interactions at the highest energies attainable in a laboratory,
overpassed only by astrophysical sources.\newline
\newline
\textbf{Acknowledgement}: The authors are grateful to their families for
support and patience. Concretely, (AB) and (SEE) would like to thank their
mothers: \textbf{Fatima} and \textbf{Fettouma}, respectively.

\end{document}